# Anharmonic Lattice Dynamics and Anisotropic Electron-Phonon Coupling in Quasi-1-Dimensional Charge Density Wave $Ta_2NiSe_7$

Prithwija Mandal,[1] S. Nanthini,[2] Aditya Singh,[1] Kewal S. Rana,[3] Dibyendu Dey,[2] Kanishka Biswas,[3] and Ajay Soni[1]*

[1]School of Physical Sciences, Indian Institute of Technology Mandi, Mandi, 175005, Himachal Pradesh, India,

[2]Department of Physics and Nanotechnology, SRM Institute of Science and Technology, Kattankulathur, 603203, Tamil Nadu, India,

[3]New Chemistry Unit, International Centre for Materials Science and School of Advanced Materials, Jawaharlal Nehru Centre for Advanced Scientific Research, Jakkur, Bangalore, 560064, Karnataka, India

**Abstract:** Microscopic origin of charge-density-wave (CDW) formation in quasi-one-dimensional (quasi-1D) $Ta_2NiSe_7$ remains actively debated, particularly regarding the relative contributions of Fermi-surface nesting (FSN), electron–phonon (*e–ph*) coupling, and lattice instabilities. Here, we combine temperature- and orientation-dependent polarized Raman spectroscopy with first-principles calculations to uncover the anisotropic electronic and lattice interactions governing the CDW state in $Ta_2NiSe_7$. Heat-capacity and electrical transport measurements identify an incommensurate CDW transition at ~ 61 K. Raman spectroscopy reveals pronounced in-plane anisotropy, with *e–ph* coupling strength along intrachain *b*-axis exceeding five times than along interchain *c*-axis, whereas lattice anharmonicity is enhanced by threefold along *c*-axis. First-principles calculations identify Ta(2)–Se octahedral vibrations and Ta–Se electronic states near Fermi level as dominant channels mediating the anisotropic *e–ph* interaction. Exceptionally strong and directional *e–ph* coupling along *b*-axis establishes lattice-driven electronic instability as primary mechanism underlying CDW modulation and highlights the dominance of intrachain interactions in strongly coupled $Ta_2NiSe_7$. Despite this strong coupling, the CDW remains incommensurate, indicating that lattice anharmonicity provides an additional degree of freedom. Enhanced anharmonicity along *c*-axis suggests that anisotropic phonon–phonon interactions reshape the free-energy landscape and contribute to stabilizing incommensurate phase. These findings reveal a cooperative interplay between anisotropic *e–ph* coupling and lattice anharmonicity in governing CDW formation in low-dimensional quantum materials.

## Introduction

Charge density wave (CDW) is a collective electronic ground state characterized by a periodic charge density modulation accompanied by an associated lattice distortion.[1-3] The formation of CDW order is governed by the interplay between crystal structure, chemical bonding, lattice dynamics and collective electron-phonon (*e-ph*) interactions.[4, 5] The understanding how crystalline instabilities evolve in the presence of structural anisotropy is crucial for elucidating the emergence of correlated electronic phases and the quantum nature of collective excitations in CDW materials. Conventional driving mechanisms behind CDW are typically attributed to a singularity in the Lindhard function arising from Fermi surface nesting (FSN) at a nesting vector ($\overrightarrow{\boldsymbol{q_n}} = \pm 2\vec{k}_F$), as described by the Peierls instability, typically realized in 1D and quasi-1D systems.[1, 6] However recent reports[2, 7] suggest FSN may not fully account for CDW formation, particularly in many two-dimensional (2D) materials where $\overrightarrow{\boldsymbol{q_n}}$ deviates from the CDW modulation wave vector ($\vec{\boldsymbol{q}}_{\boldsymbol{CDW}}$). In such cases, the CDW instability is driven either by wave-vector-dependent *e-ph* coupling,[2, 8, 9] unconventional electronic correlations,[10] anharmonicity[11, 12] or by their collective interplay. As a result, the study of CDW in such systems tends to be highly material-specific or often requires additional quantitative mechanisms to understand the dynamics of CDW.[3]

The quasi-1D ternary transition-metal chalcogenides, particularly the $Ta_2MX_y$ family (where M = Ni, Pd, Pt; X = Se, Te; and y represents the Ta/X stoichiometric ratio), have garnered significant interest for novel fundamental studies of intertwined CDW, strong structural as well as transport anisotropy, and non-trivial topological states.[13-15] For instance, $Ta_2PdSe_6$ hosts robust topological surface states with gapless, spin-momentum locked carriers and ultra-high mobility, enabling high-performance polarisation-sensitive terahertz detection.[16] In contrast, $Ta_2NiSe_5$ undergoes a transition from a zero-gap semiconductor to an excitonic insulator around 326 K,[17, 18] whereas $Ta_2NiSe_7$ remains metallic and exhibits a CDW phase transition in the temperature range of 52.9-62 K, with the transition being highly sensitive to crystal disorder. [13, 19-21] Although previous studies have shown a well-defined signature of CDW formation in $Ta_2NiSe_7$ through thermal and electrical measurements,[13, 21, 22] the mechanism underlying the formation of CDW order remains elusive. Scanning tunnelling microscopy (STM) studies by Dai *et al.*[23] revealed a rich signature of CDW phase at ~ 4.2 K, including a large energy gap ($\Delta_{CDW}$) ~ 41 meV with a modulation wavelength of ~ 2*b* (where *b* is lattice parameter), consistent with the wave vector $\vec{\boldsymbol{q}}_{\boldsymbol{CDW}}$ =(0,0.483,0) measured by Fleming *et al.*[19] The synchrotron X-ray diffraction claimed two distinct CDWs, one at $\vec{\boldsymbol{q}}_{\boldsymbol{CDW}}$ and another at $\boldsymbol{2\vec{q}_{CDW}}$, which corresponds to the transverse displacement of Ni and Se and longitudinal modulation of Ta atoms, respectively.[24] Considering its metallic electronic structure and quasi-1D nature, $Ta_2NiSe_7$ is expected to exhibit a Peierls-type CDW transition arising from FSN, where $\overrightarrow{\boldsymbol{q_n}}$ coincides with $\vec{\boldsymbol{q}}_{\boldsymbol{CDW}}$.[25, 26] However, the angle-resolved photoemission spectroscopy analysis on $Ta_2NiSe_7$ fails to find any nesting at the main CDW wavevector, $\vec{\boldsymbol{q}}_{\boldsymbol{CDW}}$ rather suggests a plausible nesting at $\boldsymbol{2\vec{q}_{CDW}}$.[20] A recent electric field effect study[27], uncovered an unconventional transport behaviour, characterised by a nearly gate-independent $T_{ICDW}$ and a reversal in sign of gate-induced resistivity change. These observations collectively suggest that the conventional weak-coupling Peierls mechanism, where the Fermi surface plays the dominant role in driving the CDW transition, may not adequately describe the CDW formation in $Ta_2NiSe_7$. Taken together, these assessments and available experimental evidence indicate that FSN is insufficient to solely drive CDW in this system.[27] To further clarify the role of FSN, it is necessary to determine whether the CDW transition occurs in the weak- or strong-coupling regime. The coupling strength can be evaluated using Bardeen-Cooper-Schrieffer (BCS) formalism through the dimensionless ratio $\frac{2\Delta_{CDW}(0)}{K_B T_{CDW}}$ (where, $T_{CDW}$ is the transition temperature, $K_B$ is the Boltzman constant), which has a value of 3.52 in weak-coupling limit and the ratio, significantly larger than this value are indicats strong-coupling behaviour. An additional criterion is provided by the ratio $\frac{\Delta_{CDW}}{\hbar\theta_D}$ (where $\theta_D$ is the Debye temperature, and ħ=h/2π where h is Planck's constant). A value close to unity marks the boundary between the weak- and strong-coupling limit. For the case of $Ta_2NiSe_7$ the $\frac{2\Delta_{CDW}(0)}{K_B T_{CDW}}$ ratio ~16 > 3.52 ( $\Delta_{CDW}$ ~ 41meV[23] and $T_{CDW}$ ~ 61 K) exceeding the BCS limit for weak coupling, and the $\frac{\Delta_{CDW}}{\hbar\theta_D}$ ratio ~ 2.2 ($\theta_D$ ~ 210K [13] ) also surpassses the weak coupling limit of 1, implying that $Ta_2NiSe_7$ resides in a strong coupling limits.[27] Hence, an alternative mechanism involving strong *e-* *ph* coupling is expected to play a crucial role in stabilizing the CDW state. However, this interpretation leads to an apparent paradox, as strongly coupled CDW systems often lead to commensurate lattice distortion, whereas $Ta_2NiSe_7$ exhibits incommensurate CDW (ICDW)[21, 27]. Reconciling these seemingly contradictory characteristics remains a central challenge in understanding the origin of the CDW. This unresolved puzzle requires the exploration of alternative factors that may govern the CDW state. In this regard, the intrinsic structural anisotropy may provide a distinct perspective.

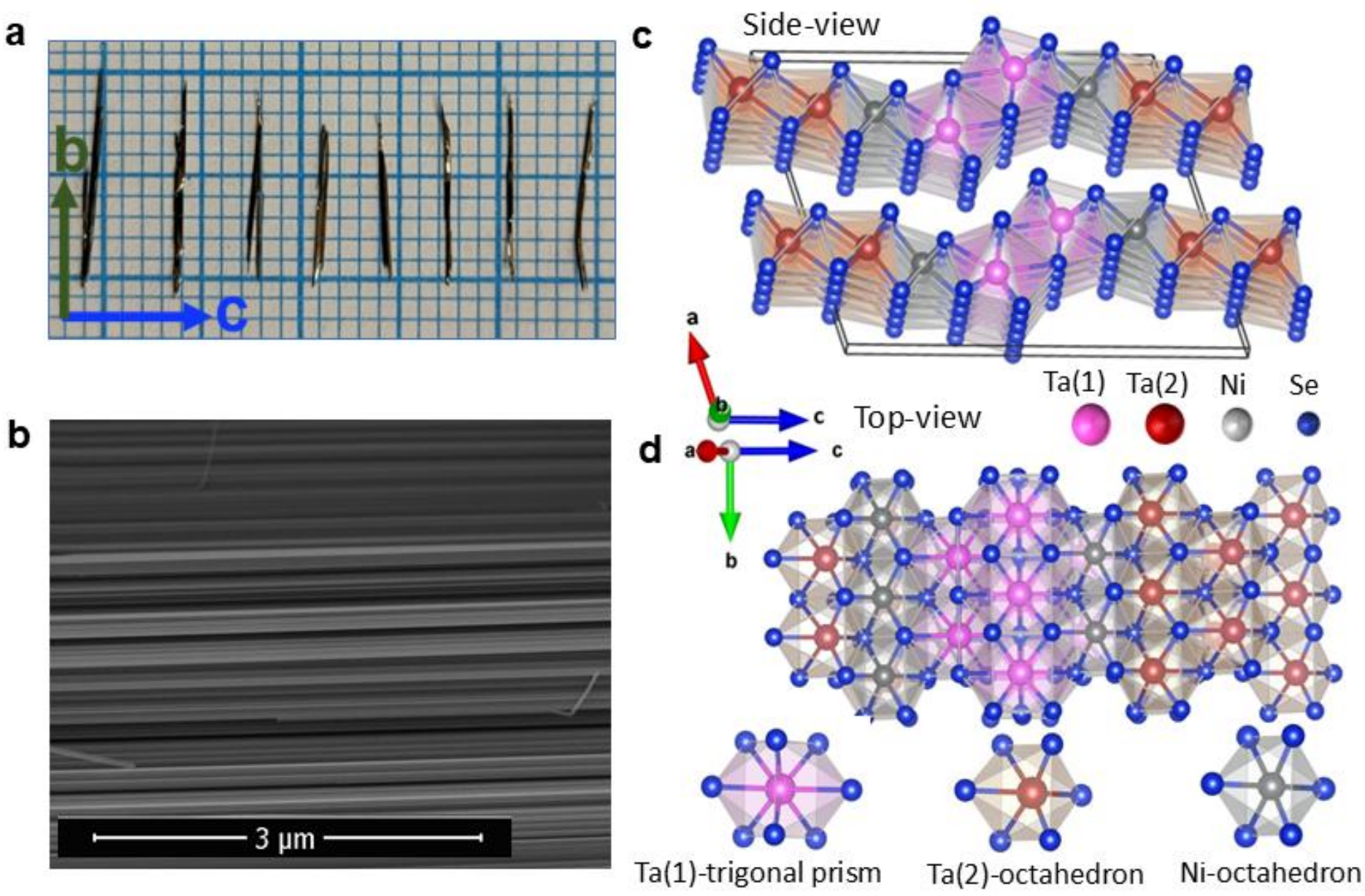

**Figure 1.** a) The photograph of the grown $Ta_2NiSe_7$ crystals, b) cross-sectional scanning electron microscopic image showing the layered nature of the crystals, c) side-view (*ac*-plane) and d) top-view (*bc*-plane) with individual polyhedra of Ta (1), Ta (2) and Ni atoms.

Quasi-1D $Ta_2NiSe_7$ exhibits an in-plane anisotropy in *bc*-plane, as evidenced by orientation-dependent polarized Raman spectroscopy, which is expected to strongly influence the phonon dynamics across $T_{ICDW}$ [28] and potentially modify the stability and periodicity of the CDW state. More broadly, chemical doping, reduced dimensionality,[29] strain engineering,[30] and high-pressure [31] studies are widely employed to tune the energetics and stability of CDW states in transition metal dichalcogenids (TMDCs) like $NbSe_2$,[32] $TaS_2$,[33] and $VSe_2$.[5] In contrast, intrinsic anisotropy may provide an alternative and largely unexplored route to directionally tune the quantum interactions.[34, 35] Herein, we clarify the effects of in-plane anisotropy on the phonon dynamics associated with CDW in $Ta_2NiSe_7$ using temperature- and orientation-dependent polarization Raman spectroscopy, supported by first-principles calculations. Our results reveal a substantial role of *e-ph* coupling in driving the CDW. The calculated Γ-centered phonon instability, together with anisotropic low-dimensional Fermi surface topology comprising electronic pockets near the Γ-point, reinforces the significance of *e-ph* coupling. Moreover, the anisotropic *e-ph* coupling, particularly enhanced along the *b*-axis, appears to favour the CDW modulation along this direction, giving rise to the pronounced zone-folding signatures observed in the corresponding orientation, while anisotropic anharmonicity is found to play an important role in stabilizing the incommensurate CDW phase. Overall, our findings reveal an intimate interplay between CDW order and crystallographic anisotropy, highlighting the potential of crystallographic directionality as an effective strategy for engineering quantum coupling parameters, thereby opening a promising avenue for understanding emergent macroscopic functionalities in low-dimensional quantum materials.

## 2. Result and Discussion

### 2.1. Structural, morphological and elemental characterization

The crystals of $Ta_2NiSe_7$ grow in an elongated needle-like (Figure 1a) shape, indicating quasi-1D growth along the crystallographic *b*-axis.[13, 21] The cross-sectional SEM image (Figure 1b) shows the layered nature of the grown crystals. The detailed structural analysis of $Ta_2NiSe_7$ has been investigated by Rietveld refinement of a powder X-ray diffraction (XRD) pattern (Supporting Information Figure S1a), which confirms a monoclinic structure (*C*2/*m* space group) with lattice parameters a ~13.86 Å, b ~3.48 Å, c ~18.63 Å, $\beta$ ~109.04º and unit cell volume ~ 852.29Å$^3$.[21, 36] The conventional unit cell is composed of coordinated polyhedra of Ta (1), Ta (2) and Ni atoms along the *b*-axis, where Ta (1) is arranged in a bicapped trigonal prismatic co-ordination with eight Se atoms, Ta (2) is arranged in an octahedral co-ordination with six Se atoms and Ni is arranged in a distorted octahedral configuration (Figure 1c, d).[20] These atoms are arranged in a unique chain structure with a staggered van der Waals (vdW) gap, causing anisotropy in the *bc*-plane and enhancing its quasi-1D nature. A similar structural in-plane anisotropy has also been observed in several 2D materials, including black phosphorus and $PdSe_2$,[37] endowing them with directional dependence of optical, electrical and thermal properties. The adjacent layers are stacked along the *a*-axis through weak vdW interactions (Figure 1c, side-view). Elemental mapping using energy dispersive X-ray spectroscopy (EDX) confirms the homogeneous distribution of elements (Supporting Information Figure S1c).

### 2.2. CDW transition temperature characterization and electronic structure:

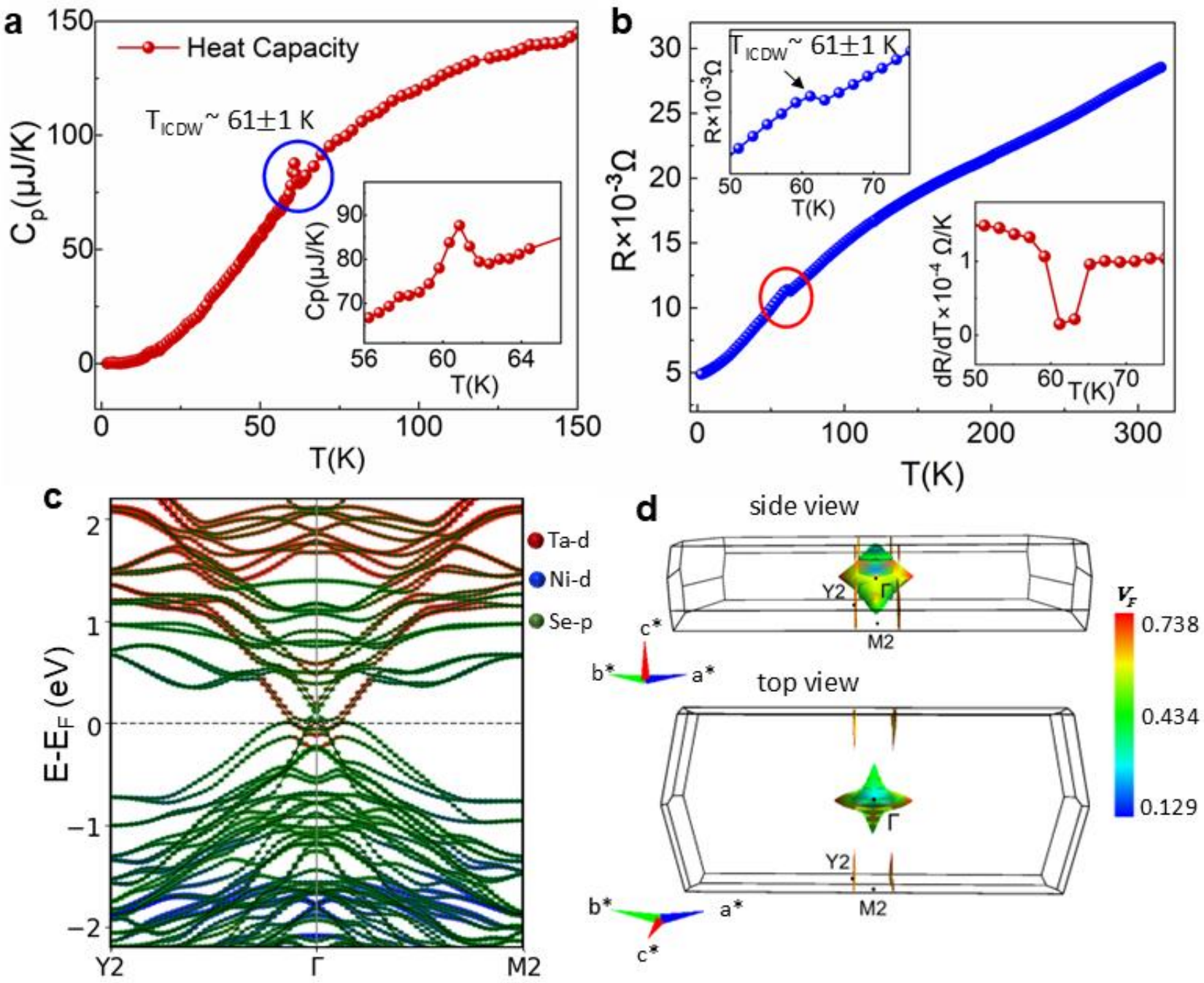

**Figure 2.** Temperature dependence of a) heat capacity $C_p(T)$ with inset showing an enlarged data at the CDW transition, b) resistance R(*T*), with top inset showing an enlarged data at the CDW transition and corresponding differential resistance in bottom inset, c) electronic band structure showing metallic nature as multiple bands are crossing the Fermi level ($E_F$), d) Side and top view of calculated Fermi

surface, showing anisotropic low-dimensional Fermi surface topology with electronic pockets centered near the Brillouin-zone center. The colour scale bar represents the magnitude of Fermi velocity.

A clear kink in $C_p(T)$ is observed at $T_{ICDW} \sim 61 \pm 1$ K, marking the onset of the ICDW transition, with an inset showing an enlarged view of the transition region (Figure 2a). The corresponding R($T$) (Figure 2b) exhibits an upturn around the same temperature (upper inset), with a dip in the dR/dT (lower inset), in agreement with previous reports and corroborating the $C_p$ results.[13] The calculated electronic band structure (Figure 2c) shows the presence of multiple bands crossing the Fermi level ($E_F$), indicating a finite electronic density of states at the $E_F$ and supporting the experimentally observed metallic transport behaviour above the $T_{ICDW}$. The orbital-projected band structure further shows that the valence band is dominated by Se-*4p* states, while the conduction band mainly originates from Ta-*5d* orbitals with additional hybridized contributions from Se-*4p* states. In contrast, Ni-derived states contribute only weakly in the vicinity of the $E_F$. Figure 2d shows the calculated Fermi surface, where the colour scale represents the magnitude of the Fermi velocity. The Fermi surface occupies a small fraction of the Brillouin zone and exhibits a pronounced anisotropic topology, which reflects the quasi-1D and metallic electronic character of $Ta_2NiSe_7$. Therefore, the dominance of the Ta-Se chain network in the electronic structure facilitates the enhancement of *e-ph* coupling along the Ta chains parallel to the *b*-axis, providing a plausible origin for the pronounced anisotropic *e-ph* coupling observed in this material system.

## 2.3 Signature of phonon instability from theoretical calculations

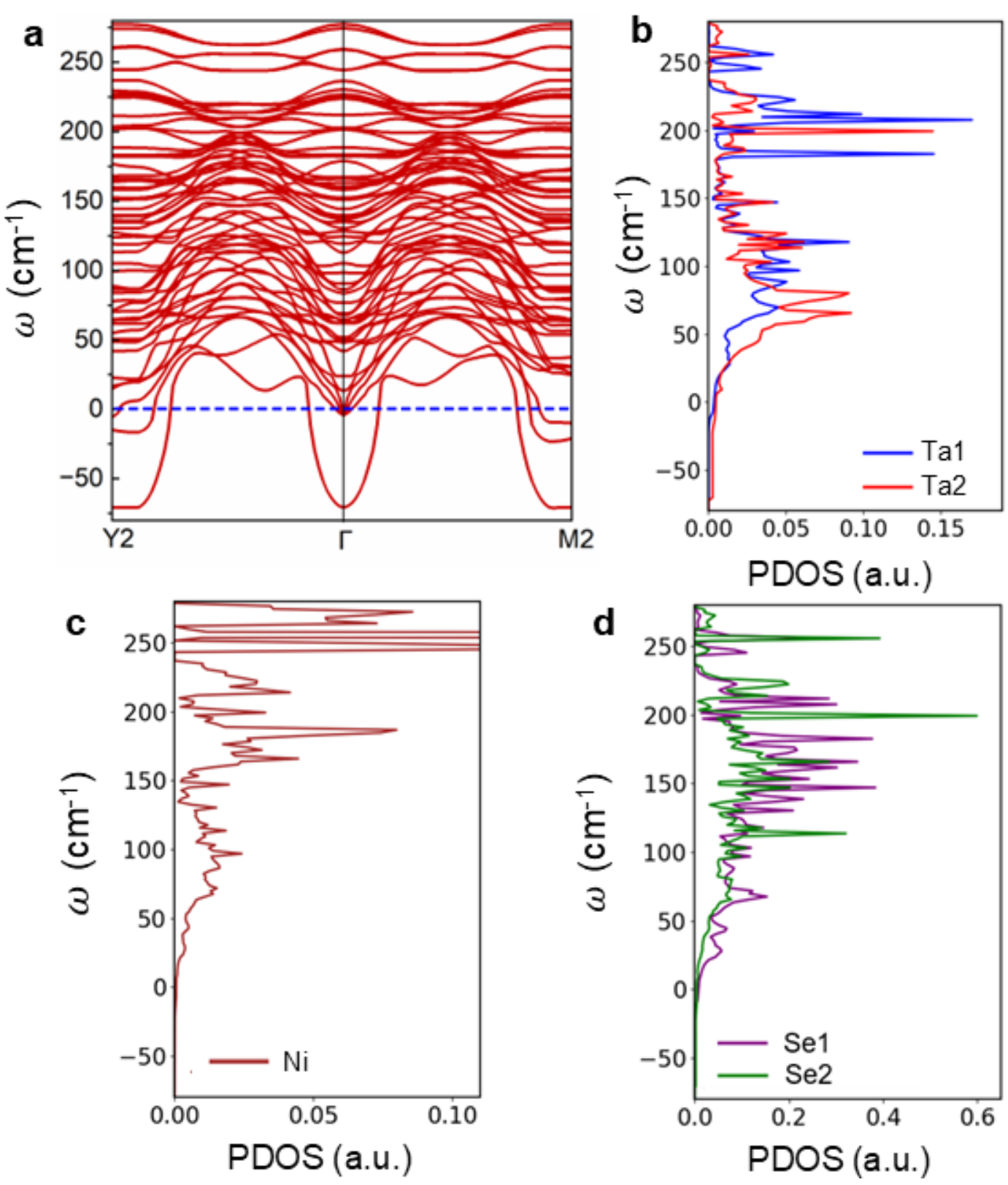

**Figure 3**. a) Calculated phonon dispersion along the high-symmetry directions ($Y_2 - \Gamma - M_2$), showing distinct low-energy imaginary phonon branches around the Γ point and at finite wave vectors approaching $q_b \sim 0.5$ along the $\Gamma - M_2$ ( $\Gamma - Y_2$) direction, close to the experimentally reported CDW

modulation vector, b–d) atom-projected phonon density of states (PDOS) for Ta, Ni and Se, respectively.

Phonon calculations show imaginary branches with instability around the Γ point as well as at finite wave vectors approaching $\vec{q}_b \sim 0.5$ along the $\Gamma$–$M_2$ ($\Gamma$–$Y_2$) direction in the phonon dispersion (Figure 3a). Notably, this finite-$q$ instability occurs close to the experimentally reported CDW modulation vector, $\vec{q}_{CDW} \approx (0,\ 0.483,\ 0)$, suggesting a direct connection between the calculated lattice instability and the observed incommensurate CDW modulation. The strong low-energy instability at Γ reaches approximately -70 $cm^{-1}$, and the symmetry analysis identifies an unstable mode as a Raman-active $B_g$ phonon (Supporting Information *Table-S1*). Such imaginary frequencies indicate that the high-symmetry monoclinic phase is dynamically unstable against specific collective lattice distortions within the harmonic approximation. The instability remains confined primarily to the selected low-energy optical branches, suggesting that the lattice instability is highly mode selective. The atom-projected phonon density of states (Figure 3b-d) shows the unstable low-frequency vibrational region, which is dominated primarily by contributions from the Ta (2)-Se sublattice, whereas the Ta (1)-Se contributions remain distributed mainly within the stable positive-frequency region. Here, the instability is strongly associated with vibrations due to the distortions involving the Ta (2)-Se octahedral chains. Hence, the phonon dispersion calculations suggest that lattice degrees of freedom actively contribute to the CDW instability. Accordingly, Raman spectroscopy was employed to investigate the lattice dynamics and *e-ph* coupling associated with the emergence of the CDW state.

2.4. Room temperature orientation-dependent polarized Raman spectroscopic charactarization.

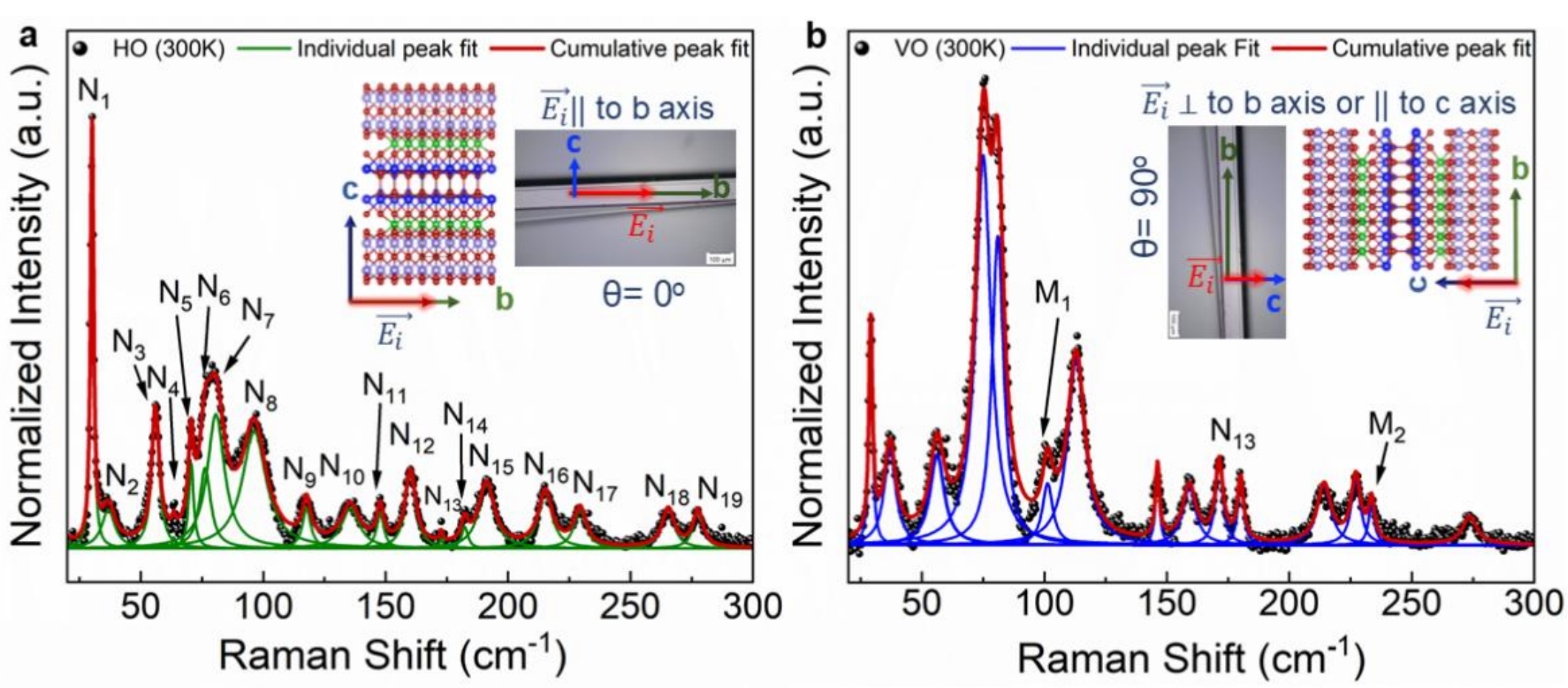

**Figure 4**. Orientation-dependent polarized Raman spectra at room temperature with the polarization vector ($\vec{E}_i$) of the incident laser in parallel with the a) *b*-axis (horizontal orientation, HO), and b) *c*-axis (vertical orientation, VO) of the crystal. The insets show the orientation of the crystal with the simulated structure, where $\theta$ is the angle between the $\vec{E}_i$ and the crystal *b*-axis.

The primitive unit cell of $Ta_2NiSe_7$ consists of thirty atoms occupying the 4i Wyckoff positions. According to the group theory analysis, the irreducible representation of the optical vibrational modes at the Brillouin zone centre of $Ta_2NiSe_7$, is represented as $\Gamma_{optical} = 20A_g + 9A_u + 10B_g + 20B_u$, where $g$ and $u$ represent Raman- and infrared-active modes, respectively.[28, 38] Besides optical characterization, orientation-dependent Raman measurement (Figure 4a-b, the inset showing the orientation geometry) has been performed by rotating the crystal in two special orientations to address the structural anisotropy in the *bc*-plane, while keeping the polarizer and analyzer parallel to each other. The coexistence of these two inequivalent polyhedra Ta (1)-Se and Ta (2)-Se, introduces substantial local structural inhomogeneity within the chain network and gives rise to distinct local bonding stiffness and unique phonon dynamics. The room temperature Raman spectra (Figure 4b) show a total of nineteen Raman modes when the polarised optical vector ($\vec{E}_i$) of the incident laser coincides with the *b*-axis (HO configuration, along intrachain direction), whereas fifteen Raman modes are detected when $\vec{E}_i$ coincides with the *c*-axis (VO configuration, along interchain direction). All the experimentally observed characteristic Raman modes and their symmetry analysis show good agreement with the phonon calculation (listed in Supporting Information *Table-S1*, and see Figure S2) and the earlier reports.[28, 38] For HO, the peaks at ~30 ($N_1$, $A_g$), ~70 ($N_5$, $A_g$), ~96 ($N_8$, $A_g$), ~135 ($N_{10}$, $A_g$), ~190 ($N_{15}$, $A_g$) and ~264 cm$^{-1}$ ($N_{18}$, $A_g$) appears prominent while the peaks at ~102 cm$^{-1}$($M_1$), ~171 cm$^{-1}$ ($N_{13}$, $A_g$), and ~235 cm$^{-1}$ ($M_2$, $A_g$) modes remains undetectable. On the other hand, for VO, the $N_{13}$, $M_1$ and $M_2$ modes become prominent, and $N_5$, $N_8$, $N_{10}$, $N_{18}$ modes disappear. Therefore, the intensity of $N_1$, $N_5$, $N_8$, $N_{10}$, and $N_{18}$ modes can act as an indicator of the *b*-axis direction.[28] Thus, orientation-dependent polarized Raman spectroscopy identifies characteristic vibrational modes of $Ta_2NiSe_7$ and enables a precise identification of crystallographic axes, thereby facilitating a robust foundation for probing the anisotropic lattice dynamics and the optical response within the *bc*-plane across the $T_{ICDW}$.

2.5. Signature of CDW formation by temperature-dependent Raman measurement and evidence of e-ph coupling.

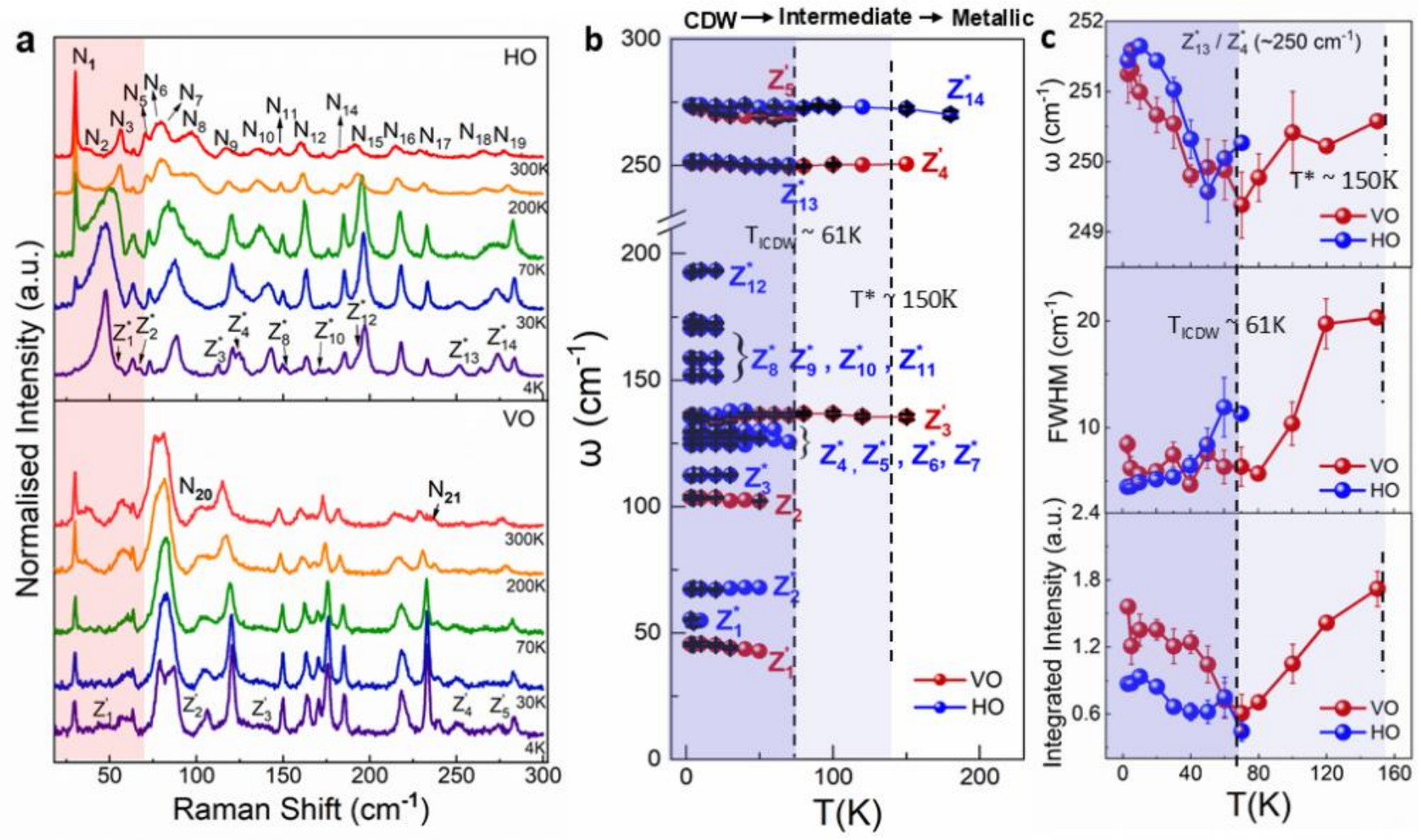

**Figure 5.** Temperature-dependent Raman spectra of $Ta_2NiSe_7$, a) in horizontal orientation (HO, upper) and vertical orientation (VO, lower) of the crystal at a few representative temperatures, and temperature evolution of b) phonon frequency of zone folded modes and, c) frequency, FWHM, and integrated intensity of the ~ 250 cm$^{-1}$ ($Z^*_{13}/Z'_4$) mode (in HO/VO).

To elucidate the effect of in-plane anisotropy on the CDW, temperature-dependent Raman measurements have been carried out in the temperature range of 3.5-300 K, in both orientations (HO and VO). A notable spectral difference has been observed between the two orientations, particularly in the low-frequency range (20-100 cm$^{-1}$), which strongly points out the anisotropy-driven characteristic of collective excitations associated with CDW.[26] As the temperature decreases, multiple new modes (Figure 5a-b and Supporting Information Figure S3a-c) appear below $T_{ICDW}$, indicating the formation of CDW superlattice.[2, 29, 39] In HO, a total of fourteen new Raman modes appear, while in VO only five new Raman modes have been detected and identified. The thermal evolution of these modes shows an anomalous behaviour that is markedly different from the other normal modes.[39] For instance, the temperature dependence of the most prominent $Z_{13}^{*}/\ Z_{4}'$ (~250 cm$^{-1}$) modes (Figure 5c) reveals a minimal redshift of (< 2 cm$^{-1}$) in ω($T$), accompanied by a gradual broadening in FWHM, a reduction in integrated intensity, and eventual disappearance above $T_{ICDW}$ (Figure 5c).[40-43] These behaviours are analogous to the characteristics of the zone-folded (*ZF*) modes. Hence, these emerging modes are expected to be *ZF* modes and are named as $Z^{*}$and $Z'$. Notably, the emerged *ZF* modes are prominent when the $\vec{E}_i$ is aligned along the crystallographic *b*-axis (HO), suggesting that the underlying zone-folding is strongly affected by structural anisotropy. Consistently, the thermal evolution of *ZF* modes exhibits pronounced anisotropic response. In HO, the $Z_{13}^{*}$ (~250 cm$^{-1}$) mode (Figure 5b) disappears near $T_{ICDW}$, and the $Z_{14}^{*}$ (~273 cm$^{-1}$) mode persists up to temperature ($T$) ~ 180 K, well above the CDW transition temperature, while in VO, the $Z_{4}'$ (~250 cm$^{-1}$) mode remains detectable upto $T$ ~ 150 K, and the $Z_{5}'$ (~273 cm$^{-1}$) mode disappears near $T_{ICDW}$. The existence of these modes in both HO and VO, above $T_{I\text{-}CDW}$, provides a signature of residual short-range CDW correlations or dynamic fluctuations persisting into the normal state, which is in agreement with the electron-diffraction reports of second-order superlattice reflections persisting up to $T$ ~ 200 K.[19] Therefore, the selective emergence of *ZF* modes together with the persistence of CDW fluctuations above $T_{ICDW}$, reveals a marked anisotropy in the *bc*-plane and significant correlations on the CDW-induced lattice reconstruction.

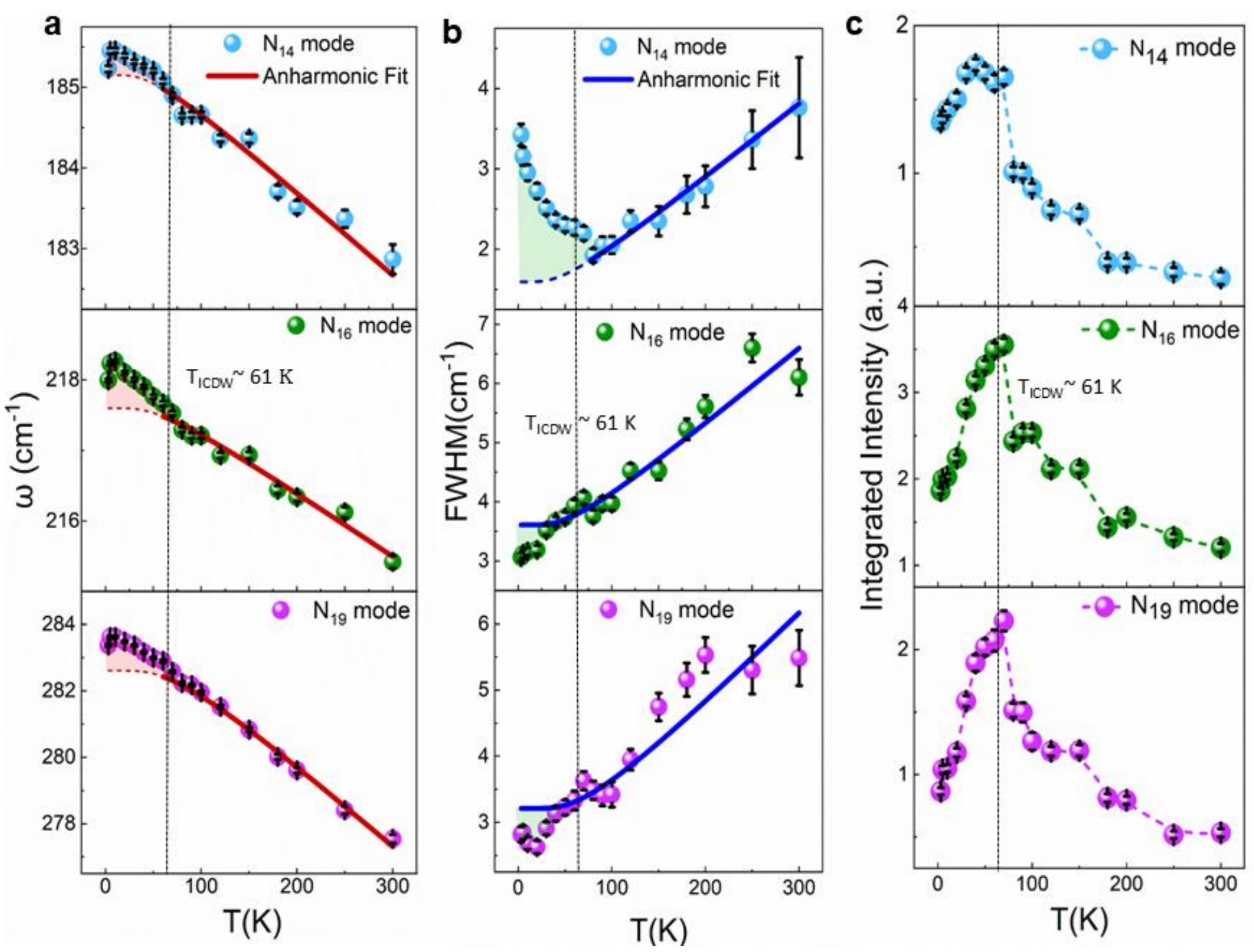

**Figure 6**. Temperature-dependent shift in a) frequency ($\omega$), b) FWHM ($\gamma$), and c) integrated intensity of the normal modes $N_{14}$, $N_{16}$, and $N_{19}$ in HO.

Figure 6a illustrates the temperature dependence of $\omega(T)$ in HO, showing a significant red shift of ~2.3 (for $N_{14}$), ~2.5 (for $N_{16}$), and ~ 5.8 cm$^{-1}$ (for $N_{19}$), with a gradual broadening in FWHM (Figure 6b). The shift in $\omega(T)$ and FWHM ($\gamma$ ($T$)) of the respective Raman modes are fitted using the anharmonic phonon-phonon (*ph-ph*) interaction model, which considers the decay of an optical phonon into two acoustic phonons.[39, 44, 45] The mathematical form of the anharmonic model for the $\omega(T)$ and $\gamma(T)$ is expressed as, $\omega(T) = \omega_0 - \delta . \left[exp\left(\frac{\hbar\omega_0}{2K_BT}\right) - 1\right]^{-1}$ and $\gamma(T) = \gamma_0 - \delta' . \left[exp\left(\frac{\hbar\omega_0}{2K_BT}\right) - 1\right]^{-1}$, where $\omega_0$ and $\gamma_0$ are the frequency and FWHM at absolute zero, $\delta$ and $\delta'$ are the anharmonic constants for frequency and FWHM and $K_B$ is the Boltzmann constant. All the fitting parameters are given in the Supporting Information *Table-S2* and *Table-S3*. Figure 6a-b (for HO) (Supporting Information Figure S4c (for VO)) shows a clear deviation of anharmonic behaviour marked as red and green areas below $T_{I\text{-}CDW}$, indicating an involvement of additional *e-ph* interaction.[39, 44, 45] In contrast, above $T_{ICDW}$, the systematic softening is accompanied by linewidth broadening, indicating enhanced thermal fluctuations and anharmonic *ph-ph* interactions, which dominate the lattice dynamics and destroy the CDW phase.[11] The temperature-dependent integrated intensity of these modes (Figure 6c) increases upon cooling and attains a maximum around $T_{ICDW}$, followed by a substantial reduction at low temperatures, which indicates a coupling of these Raman modes with CDW order parameter and ordering below $T_{ICDW}$. Similar observations are reported in different materials such as $ErTe_3$,[46] 1T-$VSe_2$,[2] suggesting that near the transition temperature, electronic spectral weight is strongly redistributed, which amplifies the effective *e-ph* scattering probability. In the CDW phase, the opening of a partial CDW gap reduces the available electronic density of states, causing the reduction in integrated intensity below $T_{I\text{-}CDW}$.[2, 47] Therefore, the (i) emergence of *ZF* modes, (ii) anomalous deviation of $\omega(T)$ and $\gamma(T)$ from intrinsic anharmonic behaviour, and (iii) enhancement of integrated intensity of the normal modes, around $T_{I\text{-}CDW}$, collectively indicate the formation of CDW superlattice and provide evidence of *e-ph* coupling to drive the CDW.

2.6. Anisotropic *e-ph* coupling and lattice anharmonicity along *b* and *c*-axis.

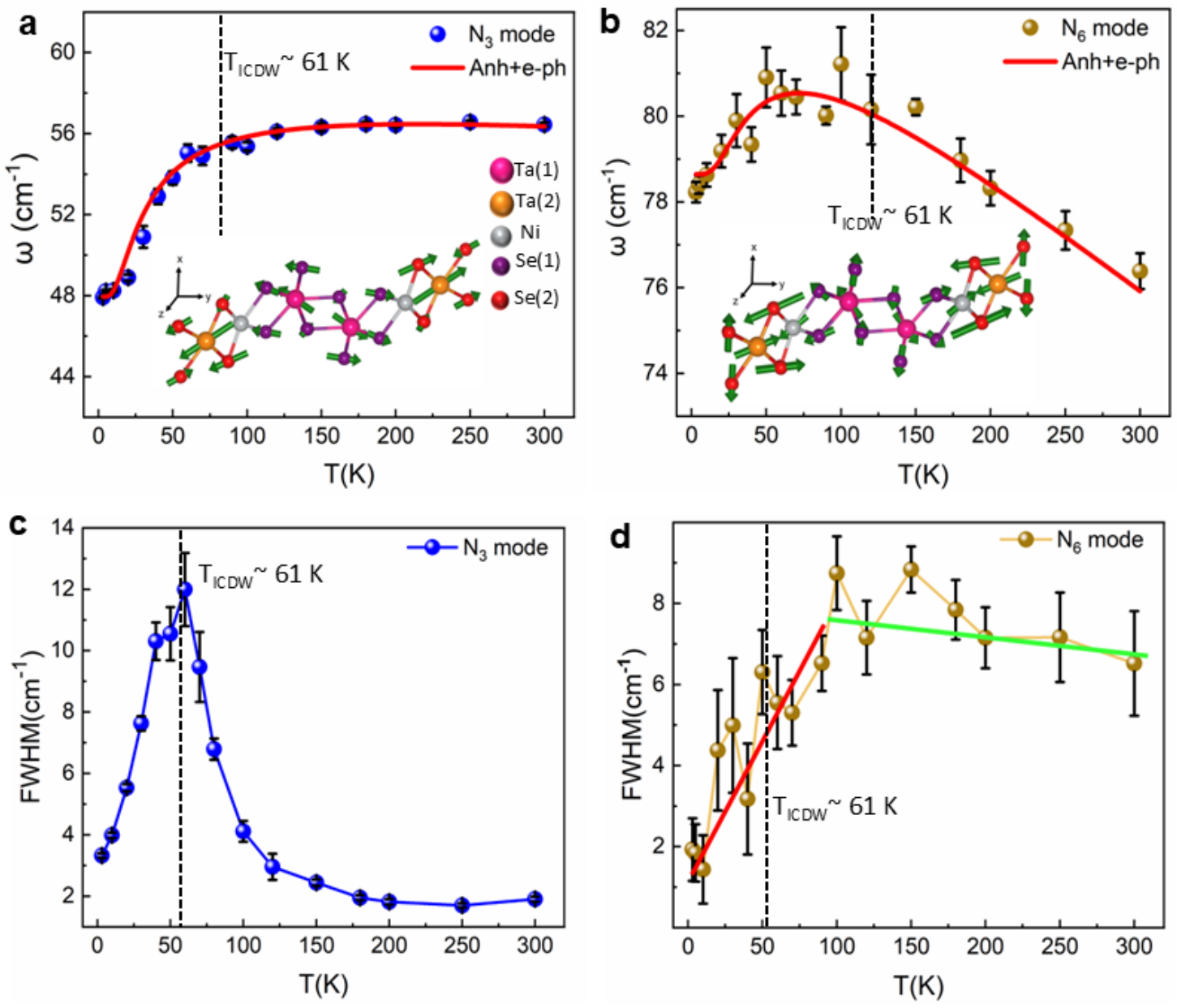

**Figure 7.** Temperature-dependent a,b) frequency shift, and c,d) FWHM shift of $N_3$ and $N_6$ modes in HO. The red and the green straight lines in Figure 7d are shown as a guide to the eye.

Apart from identifying the signature of *e-ph* coupling, the strength of *e-ph* coupling is expected to change along different axial directions. In order to evaluate the relative coupling strength with anisotropy, the $\omega(T)$ of $N_3$ and $N_6$ modes is compared for HO and VO. The insets of Figure 7a,b illustrate the eigen-displacement of the constituent atoms in the *yz* plane, associated with the respective Raman modes. In Figure 7a,b (for HO) (Supporting Information Figure S4d, (for VO)), the $\omega(T)$ of $N_3$ and $N_6$ modes show hardening with increasing temperature upto $T_{ICDW}$ (< 61±1 K) followed by softening with further increase in temperature. To comprehend the anomalous hardening of $\omega(T)$ below $T_{ICDW}$, we have considered that phonon self-energy can be renormalized through anharmonic decay as well as through the creation of an electron-hole pair mediated by *e-ph* coupling. To quantify the strength of *e-ph* coupling, the $\omega(T)$ is fitted by the following equation, $\omega(T) = \omega_0 + \Delta\omega_{anh} + \Delta\omega_{e-ph}$ , where $\Delta\omega_{anh}$ is the anharmonic contribution and $\Delta\omega_{e-ph}$ $(= -\lambda . h(T, \omega_0))$ is the *e-ph* coupling contribution. The $\lambda$ is *e-ph* coupling constant, which characterizes the phonon renormalization arising from *e-ph* interactions and $h(T, \omega_0) = \left[f\left(\frac{-\hbar\omega_0}{2}\right) - f\left(\frac{\hbar\omega_0}{2}\right)\right]$, with $f(x)$ representing the Fermi function.[39, 48] All the fitting parameters are given in the Supporting Information *Table-S2* and *Table-S3*. The extracted effective mode-resolved *e-ph* coupling constant for $N_3$ Raman mode is found to be ($\lambda$) ~ 10.44 (HO) and ~ 2.48 (VO), while the corresponding anharmonicity constant is ($\delta$) ~ 0.098 (HO) and ~ 0.29 (VO) for $N_3$ mode, which is comparable to the strongly coupled systems.[39, 49] The nearly five-fold enhancement of $\lambda$ along the *b*-axis relative to the *c*-axis provides clear evidence of highly anisotropic *e-ph* coupling within the *bc*-plane. Thus, stronger coupling strength along the intrachain direction indicates that the $\vec{q}_{CDW}$ in $Ta_2NiSe_7$ is mainly composed of a *b*-axial component and corroborates the pronounced *ZF* modes observed for HO relative to VO. In contrast, the comparatively weaker coupling along the interchain (*c*-axis) direction reflects the quasi-1D nature of $Ta_2NiSe_7$. This is consistent with the earlier anisotropic transport measurements,[13] where resistivity along both *b*- and *c*-axis ($\rho_b$ and $\rho_c$) exhibits a CDW transition at ~ 60 K, while the interchain resistivity $\rho_c$ undergoes a markedly steeper decrease at $T_{CDW}$ than the interchain $\rho_b$. Notably, the present results reveal enhanced lattice anharmonicity along *c*-axis, corresponding to the direction of the distorted interchain atomic arrangement, with similar observations in puckered lattice systems such as $PdS_2$, $PdTe_2$, where enhanced lattice anharmonicity is along the structurally distorted direction.[50] The FWHM (Figure 7c) of the $N_3$ mode exhibits a gradual increase and attains a maxima around the $T_{ICDW}$, and further reduces at higher temperatures, which indicates that the $N_3$ mode became unstable in the vicinity of the transition due to CDW modulation. The FWHM of the $N_6$ mode also follows a similar trend, but remains constant above $T_{ICDW}$ [Figure 7d].

Further, the asymmetric Fano line shape of the $N_3$ mode in HO explicitly indicates evidence of strong *e-ph* coupling. The line shape for individual peak intensity ($I_o$) of the Raman mode can be described using the Breit-Wigner-Fano (BWF) function,[51] $I_0(\omega) = I_{i0}\frac{(1+((\omega-\omega_i)/\Gamma_i q_i))^2}{1+((\omega-\omega_i)/\Gamma_i)^2}$ , where $I_{i0}$ is the intensity of the peak, $q_i$ is the asymetry parameter where $1/|q|$ characterize the coupling strength ($|q| \rightarrow \infty$ indicates a pure Lorentzian line shape and $|q| \rightarrow 0$ indicating the Fano line shape), $\Gamma_i$ is the broadening parameter and $\omega_i$ is the centring peak frequency corresponding to the Raman shift. The pronounced Fano asymmetry of the $N_3$ mode, together with its temperature evolution shown in Supporting Information Figure S5a, provides strong evidence that this mode is coupled to the electronic continuum via a strong *e-ph* coupling and the $1/|q|$ (coupling strength) value attains a maxima (~0.6) nearly to the $T_{ICDW}$, indicating a profound increase in *e-ph* coupling in the system during CDW formatmation (Supporting Information Figure S5b). Our study reveals that direction-dependent *e-ph* coupling is crucial in the quasi-1D system to host CDW state. Notably, the enhanced anharmonicity observed along *c*-axis, which reflects stronger interchain *ph-ph* interaction, suggests a possible role of lattice dynamics in stabilizing the I-CDW modulation.[52] These results identify the cooperative interplay between anisotropic electron–phonon coupling and lattice anharmonicity as the primary mechanism governing the emergence and stabilization of the incommensurate CDW state in $Ta_2NiSe_7$. Thus, establishinshng the directional lattice dynamics as a powerful route for engineering collective electronic phases in low-dimensional quantum materials.

**Conclusion:** In summary, our combined experimental and first-principles investigation uncovers a pronounced interplay between anisotropic *e–ph* coupling and anharmonic lattice dynamics in the quasi-one-dimensional charge-density-wave material $Ta_2NiSe_7$, highlighting crucial roles of these interactions in driving CDW. We identify Ta(2)–Se octahedral vibrations and Ta–Se electronic states near the Fermi level as the primary channels mediating the *e–ph* interaction, resulting in substantially stronger coupling along the intrachain (*b*-axis) direction than along the interchain (*c*-axis) direction. This anisotropic coupling provides a natural explanation for the emergence of the larger number of zone-folded phonon modes observed below $T_{(ICDW)}$ along the *b*-axis and underscores the dominant role of lattice–electronic interactions within the chains. In contrast, the weaker coupling along the *c*-axis preserves the quasi-one-dimensional character of the system. Remarkably, the enhanced anharmonicity detected along the *c*-axis highlights the importance of anisotropic phonon–phonon interactions in stabilizing the incommensurate CDW phase. Together, these findings establish that the CDW state in $Ta_2NiSe_7$ is governed not only by anisotropic *e–ph* coupling but also by direction-dependent anharmonic lattice interactions. Our work provides direct spectroscopic evidence for the cooperative roles of electronic and lattice instabilities in shaping correlated ordered states and offers a broader framework for understanding and engineering charge-ordered phases in low-dimensional quantum materials.

Supporting Information: The authors have cited additional references within the Supporting Information. [35, 36, 38, 53-62]

# Supporting Information for

# Anharmonic Lattice Dynamics and Anisotropic Electron-Phonon Coupling in Quasi-1-Dimensional Charge Density Wave $Ta_2NiSe_7$

Prithwija Mandal,[1] S. Nanthini,[2] Aditya Singh,[1] Kewal S. Rana,[3] Dibyendu Dey,[2] Kanishka Biswas,[3] and Ajay Soni[1]*

[1]School of Physical Sciences, Indian Institute of Technology Mandi, Mandi, 175005, Himachal Pradesh, India,

[2]Department of Physics and Nanotechnology, SRM Institute of Science and Technology, Kattankulathur, 603203, Tamil Nadu, India,

[3]New Chemistry Unit, International Centre for Materials Science and School of Advanced Materials, Jawaharlal Nehru Centre for Advanced Scientific Research, Jakkur, Bangalore, 560064, Karnataka, India

These supplementary information files provide additional experimental data, including X-ray diffraction (XRD), scanning electron microscopy (SEM), energy-dispersive X-ray spectroscopy (EDX), and polarized Raman spectroscopy results. It also includes comparisons with theoretical calculations and detailed discussions that complement and support the findings presented in the main manuscript.

## I. Experimental section

### A. $Ta_2NiSe_7$ crystal growth:

Single crystals of $Ta_2NiSe_7$ were grown by the chemical vapour transport (CVT) method with 5 at. % excess Selenium as transporting agent. The constituent elements Tantalum, Nickel, and Selenium (make: Alpha Aesar, > 99.9% purity) were taken with the desired stoichiometric ratio in a quartz ampule and vacuum sealed under high pressure (< $10^{-5}$ mbar). The sealed tube was placed inside a horizontal tube furnace and heated to 400°C, followed by further heating to 850°C (source zone) at 2°C/min. After maintaining a 100°C temperature gradient for 14 days, the furnace was naturally cooled to room temperature. Shiny silvery needle-like crystals were obtained at the cold end of the quartz tube.

### B. X-ray Diffraction (XRD):

To confirm the crystal growth and phase purity, X-Ray diffraction (XRD) was performed using a Rigaku SmartLab X-ray diffractometer with Cu-$K_\alpha$ radiation (wavelength ~1.54 Å) in Bragg Brentano geometry. Rietveld refinement was carried out using FULLPROF software to determine the phase purity and lattice parameters.[1]

### C. High resolution transmission electron microscopy (HR-TEM):

To confirm the single crystalline nature of the sample, HR-TEM image was captured using a FEI make Tecnai G220 S-TWIN transmission electron microscope. To prepare the sample for TEM measurement, $Ta_2NiSe_7$ crystal was dispersed in isoproply alchohol by ultra sonication and then deposited the supernatent onto carbon-coated copper grid.

### D. Morphological and elemental characterization:

Morphological studies and elemental distribution mapping on freshly cleaved crystals were carried out by field-emission scanning electron microscopy (FESEM) and energy dispersive spectroscopy (EDS) (JFEI, USA, Nova Nano SEM 450).

**E. Physical property measurement:**

Low-temperature resistance measurement (using the four-probe method) along the growth direction (*b*-axis) and heat capacity (Cp) measurements were performed in the temperature range ~ 3.5 K to ~ 150 K, using the physical property measurement system (PPMS, Quantum Design).

**F. Raman spectroscopy:**

Raman measurements were performed using a Horiba LabRAM HR Evolution spectrometer operated in a backscattering geometry. A 633 nm solid-state diode laser was used as the excitation source, and the scattered light was collected and dispersed by a Czerny–Turner monochromator equipped with an 1800 grooves/mm grating. Ultralow-frequency filters were employed to enable detection of Raman modes close to the laser line. The spectrally resolved signal was detected using a Peltier-cooled charge-coupled device (CCD) detector. Temperature-dependent Raman measurements, in a temperature window of 3.5–300 K, were carried out using a Montana Instruments-make closed-cycle cryostat. All Raman spectra were analysed by fitting with Lorentzian and Breit-Wigner-Fano (BWF) function within the experimental uncertainty, allowing extraction of the mode frequency (ω), full width at half-maximum (FWHM, Γ), peak intensity (I), and integrated intensity (area under the curve).

**G. Computational details:**

First-principles calculations based on density functional theory (DFT) were carried out using a plane-wave basis set and projector augmented-wave (PAW) potentials,[2, 3] as implemented in the Vienna *ab initio* simulation package (VASP).[4, 5] The exchange–correlation functional was treated within the generalized gradient approximation (GGA) using the Perdew-Burke-Ernzerhof (PBE) form.[6] The plane-wave energy cutoff was set to 400 eV. Brillouin zone integrations were performed using a Γ-centered k-point mesh of $4 \times 16 \times 3$. Structural relaxations were performed until the Hellmann–Feynman forces on each atom were smaller than 0.001 eV/Å. Phonon calculations were carried out on $1 \times 2 \times 1$ supercell using density functional perturbation theory (DFPT)[7] as implemented in the PHONOPY package,[8] and the phonon dispersions were obtained along the high-symmetry directions of the Brillouin zone. VASPKIT[9] was employed to post-process the VASP calculations and generate the data required for electronic band-structure and Fermi-surface analyses. The calculated Fermi surfaces were subsequently constructed and visualized using FermiSurfer[10].

## II. Powder XRD pattern, Cross-sectional SEM, EDX elemental distribution map:

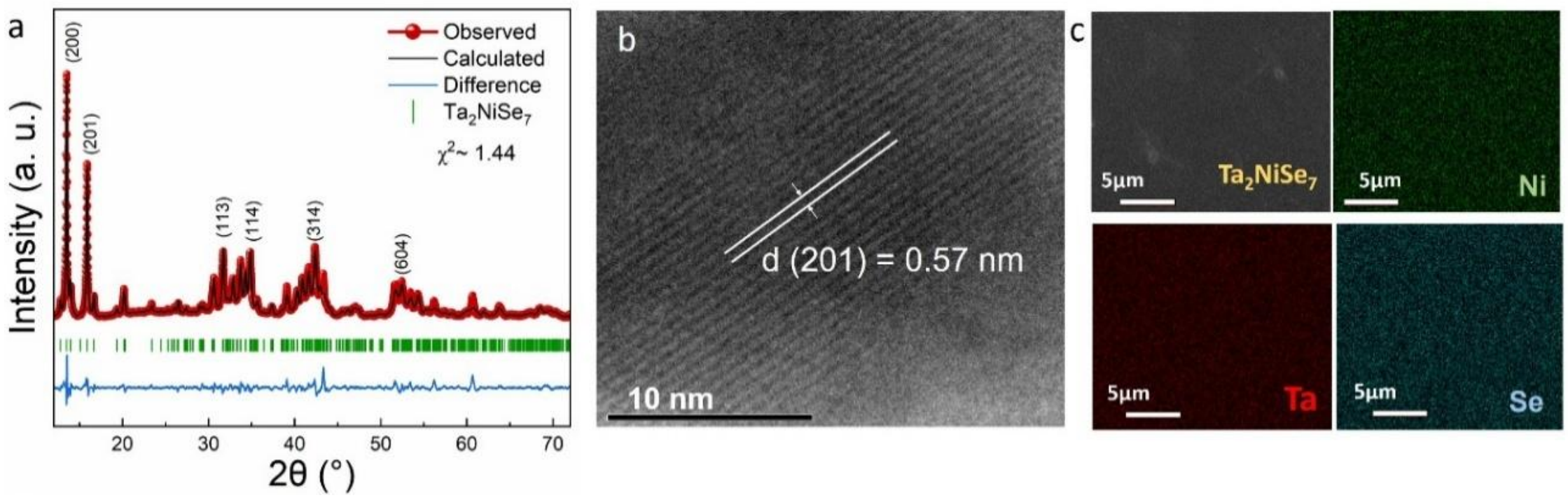

**Figure S1**. a) Rietveld-refined powder XRD of the grown $Ta_2NiSe_7$ crystals. b)High-resolution transmission electron microscopy image showing atomic spacing of (201) plane. The schematic of the monoclinic crystal structure of $Ta_2NiSe_7$, and c) the EDX elemental mapping showing the homogenius distribution of Ta/Ni/Se.

The powder XRD pattern of $Ta_2NiSe_7$ (Figure S1a) has been analysed using Rietveld refinement, confirming that $Ta_2NiSe_7$ crystallises in a monoclinic structure (C2/m, 12). The absence of any extra peak in the diffraction pattern and the fitting parameters verify the phase purity of the synthesised crystals. The high-resolution transmission electron microscopy (HR-TEM) image (Figure S1b) shows well-resolved (201) lattice planes, as separated by an interplanar distance of $d = 0.57$ nm.[11] The EDX mapping (Figure S1c) shows a homogeneous distribution of the constituent elements (Ta, Ni, and Se) throughout the sample, with no evidence of elemental agglomeration or clustering.

## III. ***Table-S1.*** **Experimentally observed Raman modes (horizontal and vertical orientation) in both ~ 3.5 K and ~ 300 K:**

| Theoretical estimation ($cm^{-1}$) | Experimental (Normal modes) at 300 K ($cm^{-1}$) | Experimental (Zone-folded modes) at 3.5 K ($cm^{-1}$) |
|---|---|---|
| -70.51 ($B_g$) | - | 45 ($Z_1'$) |
| 23.23 ($A_g$) | 30 ($N_1$) ($A_g$) | 55 ($Z_1^*$) |
| 48.13 ($B_g$) | 36 ($N_2$) ($B_g$) | 66 ($Z_2^*$) |
| 50.23 ($A_g$) | 56 ($N_3$) ($A_g$) | 103 ($Z_2'$) |
| 62.04 ($B_g$) | - | 112 ($Z_3^*$) |
| 64.54($A_g$) | 63 ($N_4$) ($A_g$) | 124 ($Z_4^*$) |
| 68.91 ($A_g$) | 70 ($N_5$) ($A_g$) | 126 ($Z_5^*$) |
| 76.65 ($A_g$) | 75 ($N_6$) ($A_g$) | 129 ($Z_6^*$) |
| 86.32 ($B_g$) | 80 ($N_7$) ($A_g$) | 135 ($Z_7^*/Z_3'$) |
| 90.46 ($A_g$) | 96 ($N_8$) (HO) ($A_g$) | 151 ($Z_8^*$) |
| 105.27($A_g$) | 102($M_1$) (VO)($A_g$) | 158($Z_9^*$) |
| 119.18 ($B_g$) | 115 ($N_9$) ($A_g$) | 170 ($Z_{10}^*$) |
| 128.38 ($A_g$) | 134($N_{10}$) (HO) ($A_g$) | 173 ($Z_{11}^*$) |
| 135.59 ($B_g$) | - | 192 ($Z_{12}^*$) |
| 144.23 ($B_g$) | 146 ($N_{11}$) ($B_g$) | 250 ($Z_{13}^*/Z_4'$) |
| 151.03 ($A_g$) | - | 273 ($Z_{14}^*/Z_5'$) |
| 158.07 ($B_g$) | - | |
| 160.2 ($A_g$) | 158 ($N_{12}$) ($A_g$) | |
| 166.21 ($A_g$) | 171($N_{13}$) ($A_g$) | |
| 178.95 ($A_g$) | 181 ($N_{14}$) ($A_g$) | |
| 182.09 ($B_g$) | - | |
| 188.52 ($A_g$) | 189($N_{15}$) (HO) ($A_g$) | |
| 201.37 ($A_g$) | - | |
| 202.70 ($A_g$) | - | |
| 223.88 ($A_g$) | 214 ($N_{16}$) ($A_g$) | |
| 224.71 ($A_g$) | 227 ($N_{17}$) ($B_g$) | |
| 230.22 ($A_g$) | 235 ($M_2$) (VO) ($A_g$) | |
| 244.43 ($B_g$) | - | |
| 259.3 ($A_g$) | 263($N_{18}$) (HO) ($A_g$) | |
| 274.25 ($A_g$) | 275 ($N_{19}$) ($A_g$) | |

*Table-S1* shows the comparison of the normal Raman modes recorded at 300 K and theoretically estimated modes. The zone-folded (ZF) Raman modes recorded at ~ 3.5K are mentioned in the third column for clarity. The normal Raman modes are labelled as "$N_n$" and "$M_n$" (here 'M' represents those modes which are only observed in VO and subscript 'n' denotes the $n^{th}$ Raman mode). The modes labelled with 'VO' or 'HO' are observed in the specific orientation, whereas those without any designation appear in both orientations. The ZF modes are marked as "$Z_n^*$" / "$Z_n'$" (HO / VO). The positions of the normal Raman modes are identified according to their frequency at ~300 K, and the ZF modes are identified according to their frequency at ~ 3.5 K. The experimentally observed symmetry of the Raman modes correctly matches the theoretical calculation, except for two modes at ~119 and ~227 $cm^{-1}$, where the background dominates.

## IV. Symmetry assessment of characteristic Raman modes using polarized Raman spectroscopy by rotating the crystal.

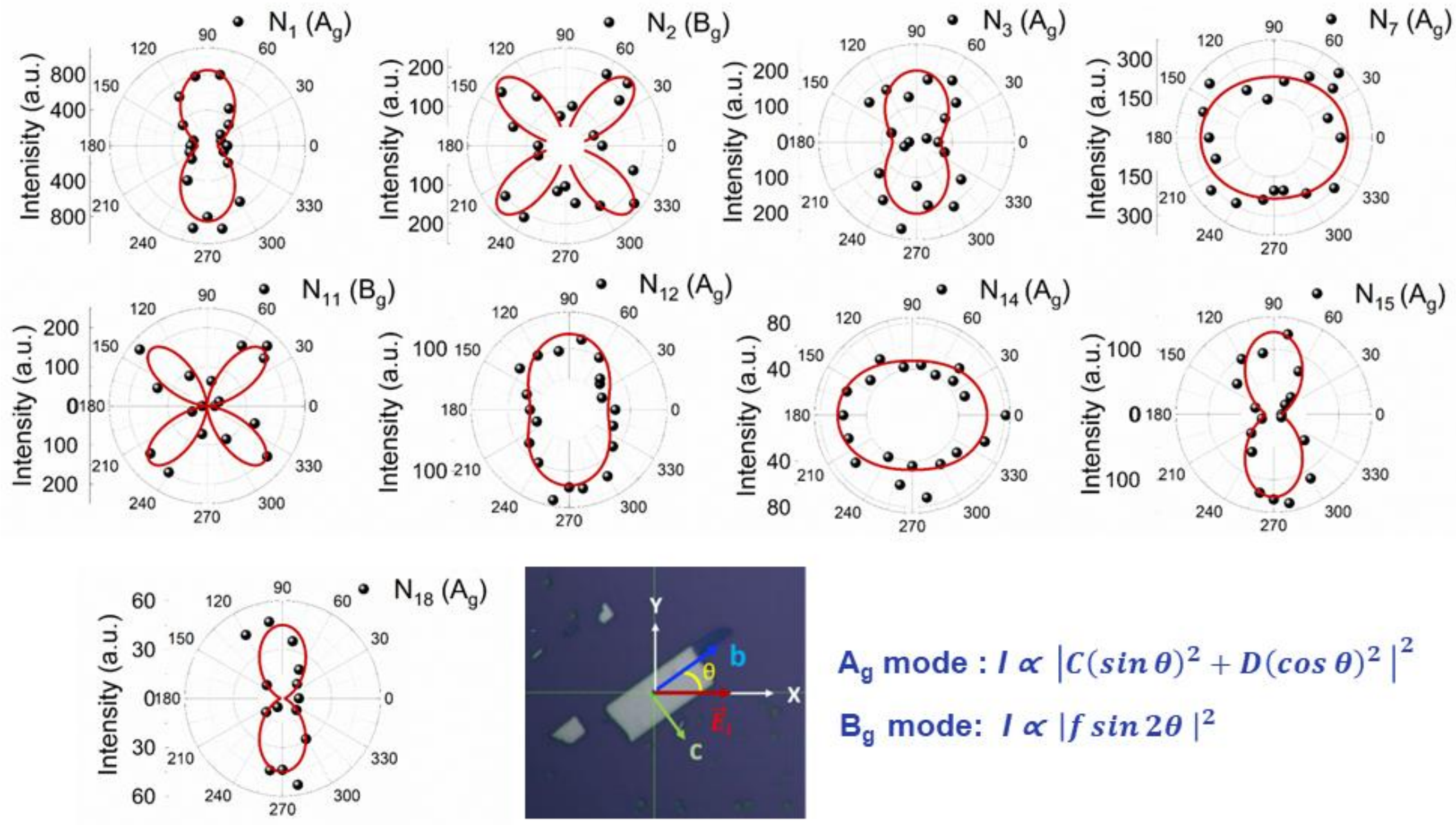

**Figure S2.** Intensities of polar plot for modes at ~ 29 $cm^{-1}$ ($N_1$), ~ 36 $cm^{-1}$ ($N_2$), ~ 56 $cm^{-1}$ ($N_3$), ~ 80 $cm^{-1}$ ($N_7$), ~ 146 $cm^{-1}$ ($N_{11}$), ~ 158 $cm^{-1}$ ($N_{12}$), ~ 181 $cm^{-1}$ ($N_{14}$), ~ 189 $cm^{-1}$ ($N_{15}$), and ~ 263 $cm^{-1}$ ($N_{18}$) with the rotation geometry of the sample. The solid red line denotes the fits to the angular dependence of the Raman intensity using equations according to the polarization selection rule for $A_g$ and $B_g$ modes.

To assign the symmetry of the Raman modes, polarized Raman measurements (Figure S2) have been carried out by keeping the polariser and analyzer fixed at θ = 0° and rotating the sample.[12, 13] The scattered intensity in the polarised Raman spectra can be expressed as, $I \propto |\overrightarrow{E_i}\,.\,R\,.\overrightarrow{E_s}\,|$, where, $\overrightarrow{E_i}$ and $\overrightarrow{E_s}$ are the unit vectors showing the direction of polarisation of the incident and the scattered light, respectively, and $R$ is the second-rank tensor. For $Ta_2NiSe_7$, we obtained R as;

$$A_g = \begin{pmatrix} a & 0 & d \\ 0 & b & 0 \\ d & 0 & c \end{pmatrix}, B_g = \begin{pmatrix} 0 & e & 0 \\ e & 0 & f \\ 0 & f & 0 \end{pmatrix}$$

Where a, b, c, d, e, and f represent the constants of the Raman tensor. The rotation geometry is shown in Figure S2, where '$\theta$' represents the rotation angle of the *b*-axis of the crystal with respect to the incident laser's polarization, $\overrightarrow{E_i}$. The intensity of the $A_g$ and $B_g$ modes shows a 'θ'-dependent evolution with the crystal rotation. To identify the vibrational symmetry, the intensity of the individual mode is fitted with the following equations (for $\overrightarrow{E_s}\,||\,\overrightarrow{E_s}$ configuration) [12, 14]

$$\text{for } A_g \text{ mode: } I \propto |C(\sin\theta)^2 + D(\cos\theta)^2|^2 \quad (1)$$

$$\text{for } B_g \text{ mode: } I \propto |f\sin 2\theta|^2 \quad (2)$$

The experimentally assigned symmetry of the Raman modes is consistent with the theoretical calculation and is listed in the following *Table-S1.*

## V. Temperature-dependent Raman spectra:

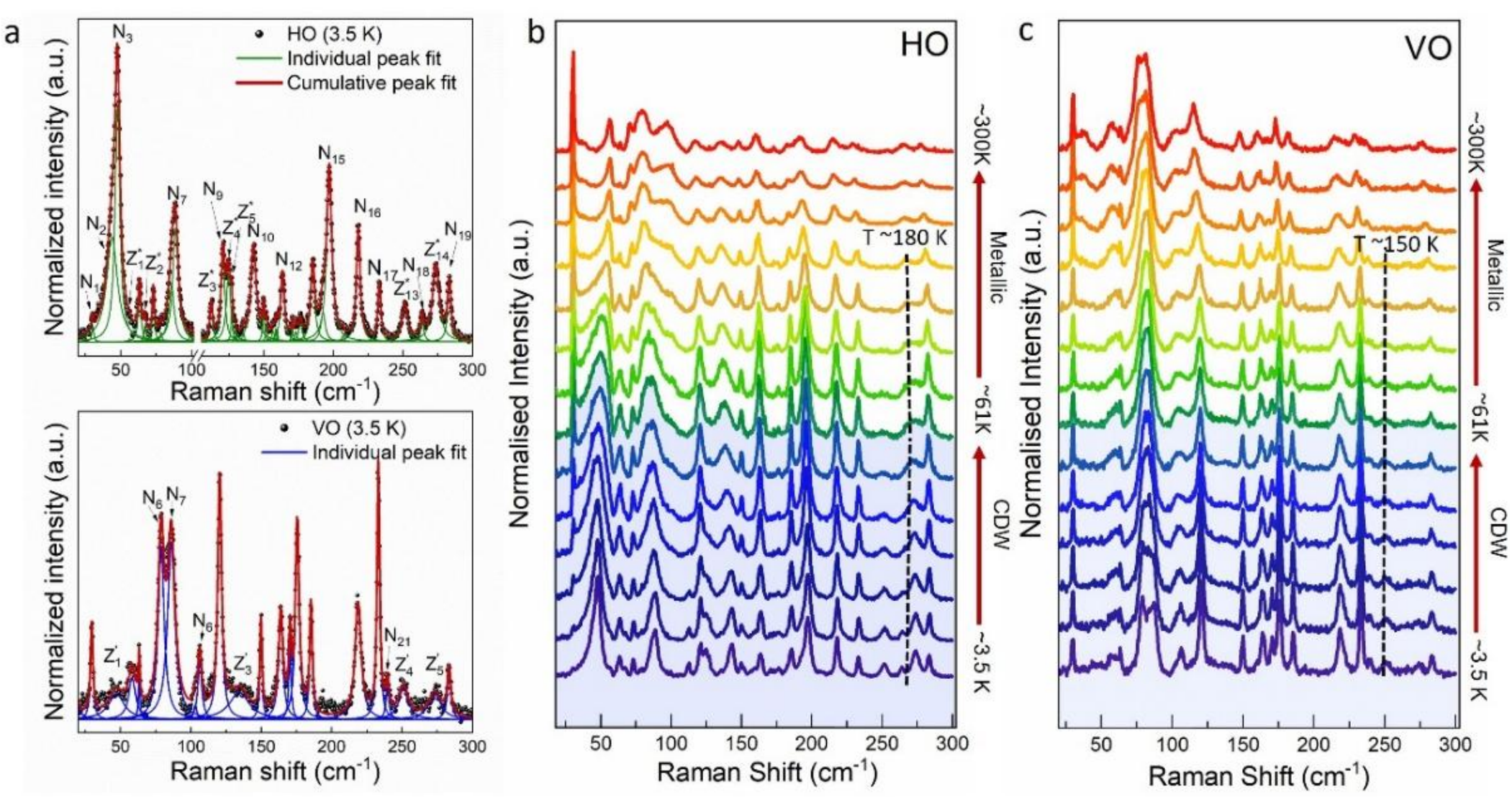

**Figure S3.** Normalized intensity vs Raman shift a) at 3.5 K, and b, c) with varying temperature (3.5 K - 300 K) for HO and VO configurations.

Figure S3a shows the distinct normal Raman modes with the CDW-induced ZF modes in the HO and VO at 3.5 K. All the identified ZF modes are listed in *Table-S1*. Figures S3b and c show the temperature dependence of Raman modes in the temperature range of 3.5-300 K.

## VI. Temperature-dependent intensity map of specific Raman modes for HO and VO. Anharmonic fitting and estimations of *e-ph* coupling in VO.

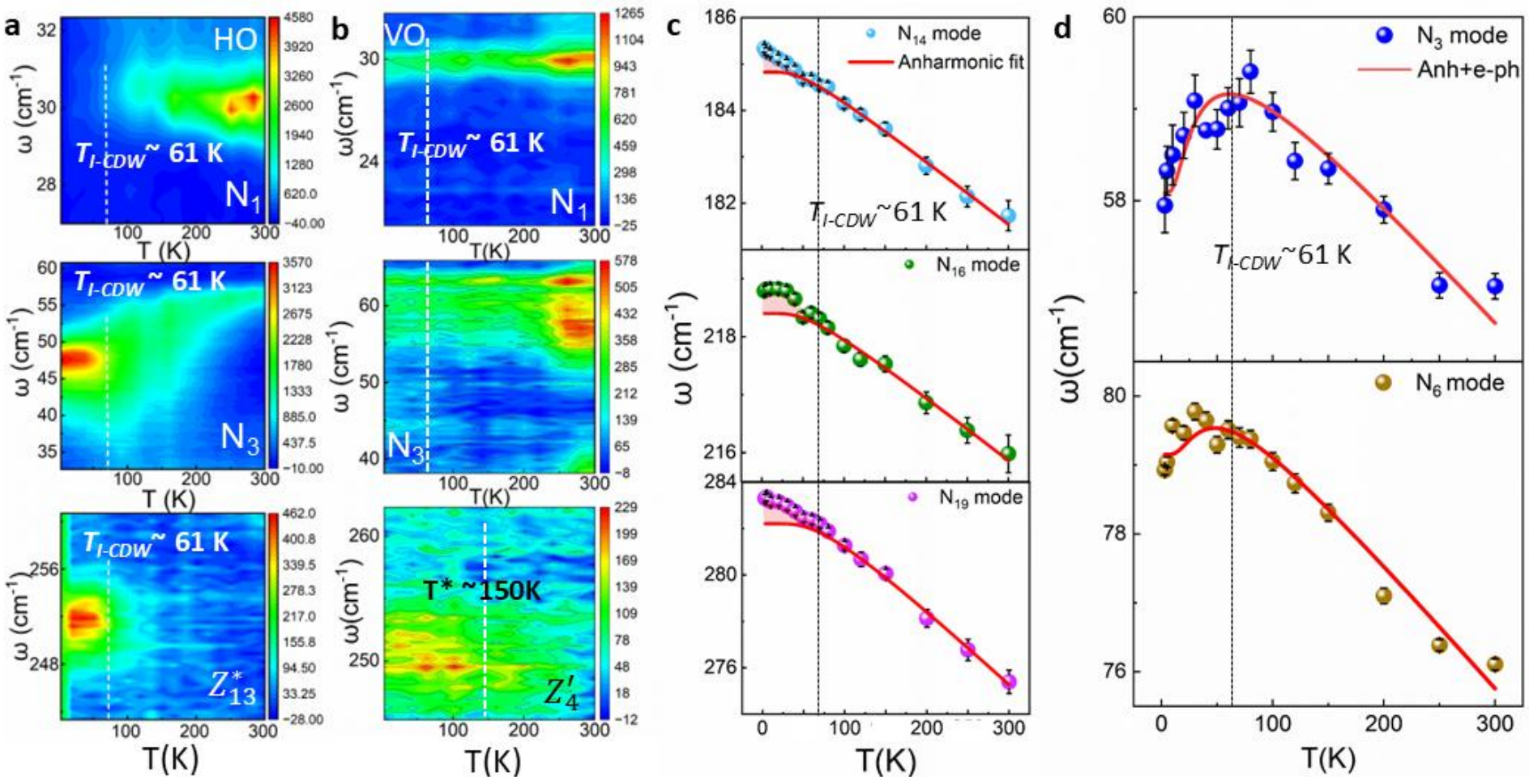

**Figure S4.** a-b) The intensity mapping color plots of the $N_1$ (~29cm$^{-1}$) (upper), $N_3$ (~56cm$^{-1}$) (middle) and $Z^*_{13}/\ Z'_4$ (~250cm$^{-1}$) (lower) Raman mode in HO and VO. Temperature-dependent c) frequency shift of $N_{14}$, $N_{16}$ , and $N_{19}$ mode fitted with anharmonic model, and d) the frequency shift of $N_3$ and $N_6$ mode fitted with anharmonic and *e-ph* coupling model for VO.

The temperature-dependent intensity mapping (Figure S4a, b) provides a comparative visualization of the spectral-weight evolution of two normal phonon modes ($N_1$ and $N_3$) and the ZF mode $Z_{13}^{*}/ Z_4'$ across the CDW transition in the HO and VO. For HO, the intensity of $N_1$ mode is very weak (almost undetectable) at temperatures below $T_{ICDW}$ and it starts to become intense when the temperature rises to $T \sim T_{ICDW}$. For VO, the same mode is detectable at low temperatures and undergoes a further enhancement with increasing temperature. The intensity map of the $Z_{13}^{*}/ Z_4'$ mode clearly shows that the persistence of the mode with increasing temperature exceeds that of $T_{ICDW}$ in VO. The contrasting intensity redistributions of both the normal and ZF phonon modes demonstrate a pronounced orientation dependence of the lattice dynamics, highlighting the strongly anisotropic optical response of the crystal and its intimate coupling to the CDW order parameter.

## VII. Analysis of Fano line shape of $N_3$ mode at varying temperature from 3.5-300K.

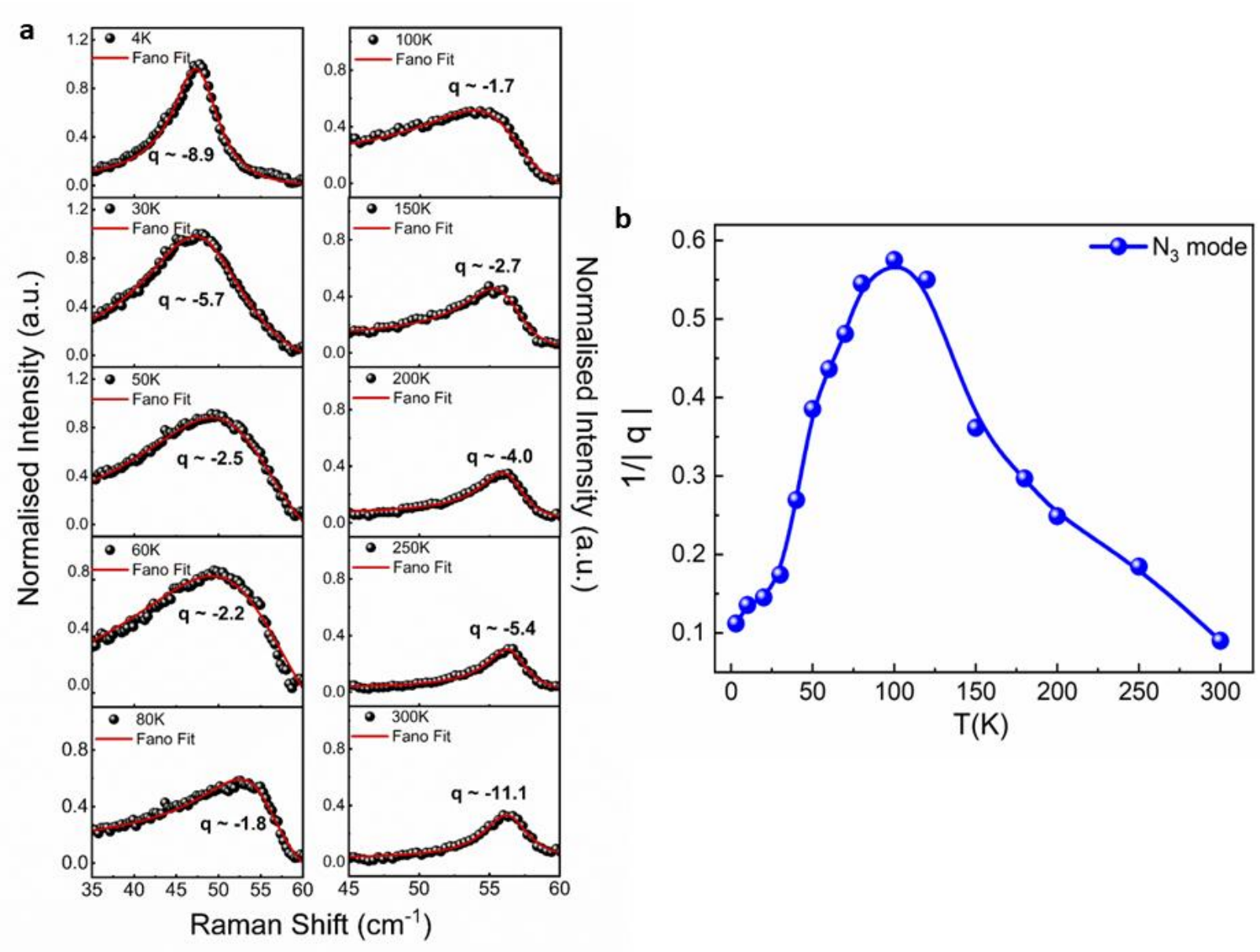

**Figure S5.** Temperature-dependent a) normalized intensity vs Raman Shift and b) 1/|q| of $N_3$ mode for HO.

For HO, the $N_3$ mode (Figure S5a) shows an asymmetric line shape, which is well fitted by the Fano function mentioned in the main manuscript. The |q| value remains high both below and above $T_{ICDW}$, while across the transition it undergoes a significant reduction, decreasing by nearly a factor of five. The temperature-dependent 1/|q| (Figure S5b) shows the variation of *e-ph* coupling across $T_{ICDW}$, highlighting a significant enhancement of charge-lattice coupling during the transition. Hence, demonstrating a critical involvement of *e-ph* coupling to drive the CDW in the system.

## VIII. Fitting parameters obtained from the ω and Γ vs T curve fitted with Anharmonic model and Anharmonic + *e-ph* coupling model in HO and VO:

***Table-S2*: For horizontal orientation (HO) of the crystal:**

| **Modes (cm$^{-1}$)** | **$\omega_o$ (cm$^{-1}$)** | **δ** | **$\Gamma_o$ (cm$^{-1}$)** | **δ'** |
|---|---|---|---|---|
| 185 ($N_{14}$) | 185.15 ± 0.057 | 1.388 ± 0.08 | 1.594 ± 0.37 | 1.241 ± 3.21 |
| 217 ($N_{16}$) | 217.597 ± 0.09 | 1.440 ± 0.18 | 3.607 ± 0.15 | 2.046 ± 1.51 |
| 283cm$^{-1}$ ($N_{19}$) | 282.605 ± 0.16 | 5.137 ± 0.44 | 3.209 ± 0.17 | 2.872 ± 2.41 |

| **Modes (cm$^{-1}$)** | **$\omega_o$ (cm$^{-1}$)** | **δ** | **λ** |
|---|---|---|---|
| 56 ($N_3$) | 58.48 ± 0.541 | 0.098 ± 0.046 | 10.44 ± 0.515 |
| 78 ($N_6$) | 84.735 ± 0.668 | 0.839 ± 0.104 | 5.25 ± 0.648 |

***Table-S3*: For vertical orientation (VO) of the crystal:**

| **Modes (cm$^{-1}$)** | **$\omega_o$ (cm$^{-1}$)** | **δ** |
|---|---|---|
| 185 ($N_{14}$) | 184.82 ± 0.082 | 1.83 ± 0.26 |
| 217($N_{16}$) | 218.4 ± 0.09 | 1.74 ± 0.37 |
| 283 ($N_{19}$) | 282.20 ± 0.20 | 6.72 ± 1.19 |

| **Modes (cm$^{-1}$)** | **$\omega_o$ (cm$^{-1}$)** | **δ** | **λ** |
|---|---|---|---|
| 56 ($N_3$) | 60.86 ± 0.274 | 0.29 ± 0.03 | 2.48 ± 0.33 |
| 78 ($N_6$) | 81.54 ± 0.379 | 0.54 ± 0.046 | 1.84 ± 0.385 |